\documentclass[aps,pra,preprint,groupedaddress]{revtex4}

\usepackage{graphicx}
\usepackage{dcolumn}
\usepackage{bm}

\begin{document}


\title{Second Order Correlation Function of a Phase Fluctuating Bose-Einstein
Condensate}


\author{L. Cacciapuoti}
\altaffiliation[Present address: ]{BNM-SYRTE, Observatoire de
Paris, 61 avenue de l'Observatoire, 75014 Paris-France; e-mail:
Luigi.Cacciapuoti@obspm.fr.}
\author{D. Hellweg}
\author{M. Kottke}
\author{T. Schulte}
\author{K. Sengstock,$^1$ W. Ertmer}
\author{J.J. Arlt}
\affiliation{Institut f\"{u}r Quantenoptik, Universit\"{a}t
Hannover, Welfengarten 1, 30167 Hannover, Germany \\
$^1$Institut f\"{u}r Laserphysik, Universit\"{a}t Hamburg, Luruper
Chaussee 149, 22761 Hamburg, Germany}
\author{L. Santos}
\author{M. Lewenstein}
\affiliation{Institut f\"{u}r Theoretische Physik, Universit\"{a}t
Hannover, Appelstra{\ss}e 2, 30167 Hannover, Germany}

\date{\today}

\begin{abstract}
The coherence properties of phase fluctuating Bose-Einstein
condensates are studied both theoretically and experimentally. We
derive a general expression for the $N$-particle correlation
function of a condensed Bose gas in a highly elongated trapping
potential. The second order correlation function is analyzed in
detail and an interferometric method to directly measure it is
discussed and experimentally implemented. Using a Bragg
diffraction interferometer, we measure intensity correlations in
the interference pattern generated by two spatially displaced
copies of a parent condensate. Our experiment demonstrates how to
characterize the second order correlation function of a highly
elongated condensate and to measure its phase coherence length.
\end{abstract}


\maketitle


\section{Introduction}
Among the various topics related to the exciting field of
Bose-Einstein condensation (BEC) \cite{Dalfovo99}, the analysis of
coherence properties of degenerate Bose gases has attracted major
interest. Coherence plays a key role in the understanding of the
fundamentals of BEC, and has a crucial importance for many
promising BEC applications, such as matter wave interferometry,
guided atomic beams, and atom lasers. The coherent character of
trapped 3D condensates well below the BEC transition temperature
$T_{c}$ has been confirmed by several experiments, using
interferometric \cite{Andrews97,Hagley99} and spectroscopic
methods \cite{Stenger99}.

However, recent theoretical and experimental developments have
shown that phase coherence is far from being an obvious property
of BEC. In particular, a phase fluctuating BEC at equilibrium has
been theoretically predicted in one-dimensional \cite{Petrov00a},
two-dimensional \cite{Petrov00,Kagan00}, and even in highly
elongated, but still three-dimensional \cite{Petrov01} trapped
Bose gases. Interestingly, in these cases the density distribution
does not differ from the usual BEC profile, since density
fluctuations are largely suppressed by the repulsive mean-field
potential. These systems are commonly called quasicondensates.
Phase fluctuations can be induced either by quantum \cite{Ho99} or
by thermal fluctuations \cite{Kane67}. For typical experimental
temperatures quantum phase fluctuations can safely be neglected as
long as the system remains in the weakly-interacting regime
\cite{Gangardt03}. The amplitude of phase fluctuations, therefore,
depends strongly on temperature and trapping geometry. In this
sense, a nearly phase coherent BEC in a highly elongated trap can
only be achieved far below $T_c$, imposing severe limitations on
experiments in constrained geometries. Phase fluctuating BECs have
been the subject of recent theoretical efforts, including the
development of a modified mean-field theory valid in all
dimensions and all temperatures below the critical point
\cite{Andersen02,Khawaja03}, the analysis of dynamic correlation
functions \cite{Luxat02}, and the extension of Bogoliubov theory
to low-dimensional degenerate Bose gases \cite{Mora02}.

The phase fluctuating nature of highly elongated BECs was first
experimentally demonstrated in Ref.~\cite{Dettmer01}. During the
ballistic expansion, phase fluctuations transform into density
modulations. The appearance of phase fluctuations and their
statistic nature were studied and the dependence of their average
value on experimental parameters was characterized
\cite{Dettmer01,Hellweg01}. Moreover, the results obtained from
measurements of the energy released during the expansion confirmed
the absence of density fluctuations in the trapped cloud
\cite{Kreutzmann03,Richard03}. Recently, the physics of
quasicondensates has been studied by means of Bragg spectroscopy,
showing that the existence of phase fluctuations leads to an
observable broadening of the momentum distribution
\cite{Gerbier02a,Richard03}. A further experiment has analyzed the
phase coherence length of non-equilibrium BECs by means of a
condensate-focusing technique \cite{Shvarchuck02}.

In this paper, we present the theoretical foundation of our
studies on coherence properties of phase fluctuating condensates.
We analyze the behavior of the second order correlation function
for our experimental conditions and provide a detailed discussion
of the experimental technique used in Ref.~\cite{Hellweg03} to
measure it. This technique is based on the analysis of the density
correlations in the interference pattern generated by a matter
wave Bragg interferometer. In analogy to the original
Hanbury-Brown and Twiss experiment \cite{Hanbury56,Hanbury56a},
our method is used to extract the phase coherence length of the
degenerate Bose gas from density correlation measurements.

This paper is organized as follows: In Sec.~II, we briefly review
the theory of phase fluctuating Bose-Einstein condensates in 3D
elongated traps \cite{Petrov01} and analyze the evolution of the
phase pattern during the ballistic expansion. The knowledge of the
free dynamics of the phase is important to closely model the BEC
evolution during the measurement process. In Sec.~III, we study
the coherence properties of the condensate and derive a general
expression for the $N$-particle correlation function of highly
elongated 3D BECs. In Sec.~IV, the experimental technique used to
measure the second-order correlation function and the phase
coherence length of the condensate is reviewed in detail.
%
\section{Phase fluctuating condensates}
In this section, we present the phase operator of a highly
elongated condensate \cite{Petrov01} and develop an analytic
description of the ballistic expansion of the fluctuating phase
pattern. These results, when combined with the free evolution of
density modulations presented in \cite{Dettmer01,Hellweg01},
provide a full understanding of the order parameter dynamics
during the time-of-flight.
\subsection{Phase operator}
In the following, we consider a cylindrically symmetric condensate
in the Thomas-Fermi regime, where the repulsive mean-field
interaction exceeds the radial ($\hbar \omega_{\rho}$) and the
axial ($\hbar \omega_z$) trap energies. At $T=0$, the density
profile has the well-known shape $n_0(\rho,z)=n_{0\mbox{\small m}}
(1-\rho^2/R^2-z^2/L^2)$, where $n_{0\mbox{\small m}}=\mu/g$
denotes the maximum density of the condensate, $\mu$ is the
chemical potential, $g=4\pi\hbar^2a/m$ the interaction constant,
$m$ the atomic mass, and $a>0$ the scattering length. Under the
condition $\omega_{\rho}\gg\omega_z$, the radial size of the
condensate, given by the Thomas-Fermi radius
$R=(2\mu/m\omega_{\rho}^2)^{1/2}$, is much smaller  than the axial
size, which corresponds to the Thomas-Fermi length
$L=(2\mu/m\omega_z^2)^{1/2}$.

Due to the repulsive mean-field energy, density fluctuations are
strongly suppressed in a trapped BEC. Therefore, the field
operator describing the condensate can be written in the form
$\hat\psi({\bf r})=\sqrt{n_0({\bf r})}\exp(i\hat\phi({\bf r}))$,
where the phase operator is defined by (see e.g.
Ref.~\cite{Shevchenko92})
\begin{equation}
\hat\phi({\bf r})=[4n_0({\bf
r})]^{-1/2}\sum_{j=1}^{\infty}f_j^{+}({\bf r})\hat a_j
+\mbox{h.c.}. \label{operphi}
\end{equation}
Here $\hat{a}_j$ represents the annihilation operator of the
quasiparticle excitation with quantum number $j$ and energy
$\epsilon_j$; $f_j^{+}= u_j + v_j$ is the sum of the excitation
wavefunctions $u_j$ and $v_j$, obtained from the corresponding
Bogoliubov-de Gennes (BdG) equations. The low-energy axial modes,
which are responsible for the long wavelength axial phase
fluctuations, have the energy spectrum
$\epsilon_j=\hbar\omega_z\sqrt{j(j+3)/4}$ \cite{Stringari98}. The
wavefunctions $f_j^+$ of these quasiparticle modes have the form
\cite{Petrov01}
\begin{equation}
f_j^{+}({\bf r})=\sqrt{\frac{(j+2)(2j+3)gn_0({\bf r})} {4\pi
(j+1)R^2 L\epsilon_j}} P_j^{(1,1)}\left(\frac{z}{L}\right),
\label{fpm}
\end{equation}
where $P_j^{(1,1)}$ are Jacobi polynomials.
Equations~(\ref{operphi}) and (\ref{fpm}) show that the phase
operator only depends on the axial coordinate $z$. In sec.~III, we
analyze the coherence properties of the condensate by studying the
correlation functions of the operator $\hat\psi({\bf r})$.
\subsection{Evolution of the phase fluctuating pattern}
Starting from the results presented in
Refs.~\cite{Dettmer01,Hellweg01}, we analyze the evolution of
phase fluctuations during the free expansion of the degenerate
Bose gas. Since the trap is highly elongated, we can assume the
condensate as an infinite cylinder, and use the local density
approximation. The time-of-flight dynamics of the order parameter
is described by the scaling law \cite{Kagan96,Castin96}
\begin{equation}
\psi(\rho,z,t)= \frac{\kappa(\tilde{\rho},z,t)}{\lambda_\rho(t)}
e^{i\frac{m\dot\lambda_\rho}{2\hbar\lambda_\rho}\rho^2}
e^{-i\frac{\mu\tilde{t}}{\hbar}}, \label{sca}
\end{equation}
where $(m\dot\lambda_\rho/2\hbar\lambda_\rho)\rho^2$ is the
quadratic phase associated with the expansion dynamics,
$\lambda_\rho^2(t)=1+\omega_\rho^2 t^2$ is the scaling
coefficient, $\tilde{t}=\int^t dt'/\lambda_{\rho}(t')^2$ is the
re-scaled time, and $\tilde{\rho}=\rho/\lambda_\rho(t)$ is the
re-scaled radial coordinate. Let $\kappa_0=\sqrt{n_0}$ be the
solution of the following equation
\begin{equation}
\left [ -\frac{\hbar^2}{2m}\nabla_{\tilde{\rho}}^2+
\frac{m\omega_\rho^2}{2}\tilde{\rho}^2 +g|\kappa_0|^2 -
\mu\right]\kappa_0=0.
\end{equation}
If we define $\kappa=\sqrt{n}\exp(i\phi)$, with $n=n_0+\delta n$,
and substitute the scaling law of Eq.~(\ref{sca}) into the
corresponding Gross-Pitaevskii equation (GPE), after linearizing
in $\delta n$ and $\phi$ we obtain:
\begin{eqnarray}
\frac{\partial (\delta n)}{\partial t}&=&
\frac{\hat\xi\phi}{\lambda_\rho^2(t)}-
\frac{\hbar}{m}\frac{\partial^2}{\partial z^2}(n_0\phi),
\label{BdG1} \\
\frac{\partial(n_0\phi)}{\partial t}&=&-\frac{\hat\xi(\delta
n/n_0)}{4\lambda_\rho^2(t)}+
\frac{\hbar}{4m}\frac{\partial^2}{\partial z^2}(\delta n) -
\frac{gn_0}{\hbar \lambda_\rho^2(t)}(\delta n), \label{BdG2}
\end{eqnarray}
where $\hat\xi=-(\hbar/m)[ n_0 \nabla_{\tilde{\rho}}^2
+\nabla_{\tilde{\rho}}n_0\nabla_{\tilde{\rho}}]$. The first term
on the right hand side of Eq.~(\ref{BdG2}) can be neglected in the
Thomas-Fermi regime. Following Ref.~\cite{Stringari98}, we average
over the radial coordinates. Let $n_I$ be the radially-integrated
unperturbed density, and $\delta n_I$ the radially-integrated
density fluctuations. From Eq.~(\ref{BdG2}) we obtain:
\begin{eqnarray}
\phi(\tilde{z},\tau)&=&\phi(\tilde{z},0)+\frac{1}{8\lambda^2\zeta}
\frac{\partial^2}{\partial \tilde{z}^2} \left [ \int_0^\tau
\frac{\delta n_I(\tilde{z},\tau')} {n_I(\tilde{z},\tau')} d\tau' \right ] \nonumber \\
&-& \frac{\zeta}{2}\int_0^\tau \frac{1}{\lambda_\rho^2(\tau')}
\frac{\delta n_I(\tilde{z},\tau')}{n_I(\tilde{z},\tau')} d\tau',
\label{phit}
\end{eqnarray}
with $\tau=\omega_\rho t$, $\tilde{z}=z/L$,
$\zeta=\mu/\hbar\omega_\rho$, and $\lambda=\omega_\rho/\omega_z$.
Equation~(\ref{phit}) can be evaluated from the known expression
\cite{Dettmer01}
\begin{equation}
\frac{\delta n_I(\tilde{z},\tau)}{n_I(\tilde{z},\tau)}=\sum_j c_j
P_j^{(1,1)}(\tilde{z})\sin\left(
\frac{a_j\tau}{1-\tilde{z}^2}\right)\tau^{-b_j},
\label{dn}
\end{equation}
where $b_j=(\epsilon_j/\hbar\omega_\rho)^2$, $a_j=b_j/\zeta$ and
\begin{equation}
c_j=\left [ \frac{(j+2)(2j+3)g}{4\pi R^2 L \epsilon_j (j+1)}
\right ]^{1/2} \frac{(\alpha_j+\alpha_j^{\ast})}{2}.
\label{ccoeff}
\end{equation}
$\alpha_j$ and $\alpha_j^{\ast}$ are random variables with a zero
mean value and $\langle |\alpha_j|^2 \rangle =N_j$, $N_j$ being
the occupation of the quasiparticle mode $j$. Near the trap
center, $\delta n_I/n_I\simeq\sum_j c_j P_j^{(1,1)}(\tilde{z})
\sin (a_j\tau)\tau^{-b_j}$, and hence
\begin{eqnarray}
\phi(\tilde{z},\tau)=&&\phi(\tilde{z},0) \nonumber \\
&&+\sum_j c_j \left \{
\frac{(j+3)(j+4)}{32\zeta\lambda^2}P_{j-2}^{(3,3)}(\tilde{z})
\int_0^\tau d\tau' \sin(a_j\tau')(\tau')^{-b_j} \right\delimiter 0 \nonumber \\
&&- \left\delimiter 0 \frac{\zeta}{2}P_j^{(1,1)}(\tilde{z})
\int_0^\tau d\tau' \frac{\sin(a_j\tau')(\tau')^{-b_j}}{1+\tau'^2}
\right \}.
\label{fullphase}
\end{eqnarray}
For large $\lambda$ and sufficiently short times-of-flight, the
significant contribution to the phase fluctuations is due to the
modes $j$ such that $\tau<<\lambda^2\zeta/[j(j+3)/4]$, and
$b_j=j(j+3)/4\lambda^2 \ll 1$. Then, using Eq.~(\ref{operphi}) for
$\phi(\tilde{z},0)$, we obtain:
\begin{equation}
\phi(\tilde{z},\tau)\simeq\sum_j c_j \left
\{1-\frac{1}{2}\arctan(\tau) \frac{j(j+3)}{4\lambda^2} \right
\}P_j^{(1,1)}(\tilde{z}).
\end{equation}
The second term in the brackets is the correction to the phase
contribution of the $j$-th mode due to the ballistic expansion.
For typical times-of-flight (tens of milliseconds), this
correction term is very small ($\simeq10^{-5}$) and the phase
pattern can be assumed as completely frozen. Using
Eq.~(\ref{fullphase}), we have verified that, for our typical
experimental parameters (see Sec.~IV-C), the phase change due to
the free evolution of the condensate is less than $\pi/10$.
%
\section{Correlation functions of a phase fluctuating condensate}
The coherence properties of a condensate are described by the
correlation functions of the field operator $\hat{\psi}$. The
importance of correlation functions becomes clear if we consider
that most experimental signals can be modelled by using this
formalism. For example, the first and second order correlation
functions, describing the single-particle and two-particle
correlation properties of the system, are connected to the
visibility of fringes in an interference experiment and to the
two-body collision rate in the condensate, respectively.

As discussed in Ref.~\cite{Petrov01}, the single-particle
correlation function of a highly elongated degenerate Bose gas can
be expressed in terms of the mean square fluctuations of the
phase:
\begin{equation}
\langle\hat{\psi}^\dagger({\bf r}_1)\hat{\psi}({\bf r}_2)\rangle=
\sqrt{n_0({\bf r}_1)n_0({\bf r}_2)}
\exp{\{-\langle[\delta\hat{\phi}({\bf r}_1,{\bf
r}_2)]^2\rangle/2\}},
\label{C_Eq_1}
\end{equation}
where $\delta\hat{\phi}({\bf r}_1,{\bf r}_2)=\hat{\phi}({\bf
r}_1)- \hat{\phi}({\bf r}_2)$ depends directly on the phase
operator $\hat{\phi}$ given in Eqs.~(\ref{operphi}). At
equilibrium, the population of the $j$-th quasiparticle mode,
$\langle\hat{a}_j^\dagger\hat{a}_j\rangle$, is a random variable
with mean value $N_j$, given by the Bose-Einstein distribution
function. The appearance of phase fluctuations is a stochastic
process governed by the temperature $T$ of the system. Since
individual realizations are not predictable, we average over an
ensemble of identically prepared condensates in thermal
equilibrium at temperature $T$. This average is indicated by
$\langle\dots\rangle_T$. When $k_BT\gg\hbar\omega_z$ ($k_B$ is the
Boltzmann constant), the population of the $j$-th mode is
$N_j\simeq k_BT/\epsilon_j$, and the thermal average of the mean
square fluctuations of the phase becomes
\begin{equation}
\langle[\delta\hat{\phi}(z_1,z_2)]^2\rangle_T=
\delta_L^2(T)f^{(1)}(z_1/L,z_2/L),
\label{C_Eq_2}
\end{equation}
where
\begin{equation}
\delta_L^2(T)=\frac{32\mu k_BT}{15N_0(\hbar\omega_z)^2}
\label{C_Eq_3}
\end{equation}
and
\begin{equation}
f^{(1)}(z_1/L,z_2/L)=
\frac{1}{8}\sum_{j=1}^{\infty}\frac{(j+2)(2j+3)}
{j(j+1)(j+3)}\left[P_j^{(1,1)}\left(\frac{z_1}{L}\right)-
P_j^{(1,1)}\left(\frac{z_2}{L}\right)\right]^2, \label{C_Eq_4}
\end{equation}
$N_0$ indicating the number of atoms in the condensate fraction.
The first order correlation function of the degenerate Bose gas is
defined by (see e.g. \cite{Scully97})
\begin{equation}
g^{(1)}_T({\bf r}_1,{\bf r}_2)=
\frac{\langle\hat{\psi}^\dagger({\bf r}_1)\hat{\psi}({\bf
r}_2)\rangle_T} {(\langle\hat{\psi}^\dagger({\bf
r}_1)\hat{\psi}({\bf r}_1)\rangle_T \langle\hat{\psi}^\dagger({\bf
r}_2) \hat{\psi}({\bf r}_2)\rangle_T)^{1/2}}.
\label{C_Eq_6}
\end{equation}
According to Eqs.~(\ref{C_Eq_1}) and (\ref{C_Eq_2}), this results
in
\begin{equation}
g^{(1)}_T(z_1,z_2)=
\exp\{-\delta_L^2(T)f^{(1)}(z_1/L,z_2/L)/2\}.
\label{C_Eq_7}
\end{equation}
For $|z_1|,|z_2|\ll L$, using the asymptotic expression of the
Jacobi Polynomials \cite{Gradsteyn65}, and summing over the
different modes in the continuous limit, one obtains an
approximated formula for the $f^{(1)}$ function valid around the
center of the condensate \cite{Petrov01}:
\begin{equation}
f^{(1)}(z_1/L,z_2/L)=|z_1-z_2|/L.
\label{C_Eq_5}
\end{equation}
In that case,
\begin{equation}
g^{(1)}_T(z_1,z_2)=
\exp\{-\delta_L^2(T)|z_1-z_2|/2L\}.
\label{C_Eq_8}
\end{equation}
This result suggests the introduction of the phase coherence
length of the condensate
\begin{equation}
L_{\phi}=\frac{L}{\delta_L^2(T)}, \label{C_Eq_9}
\end{equation}
defined as the distance at which the first order correlation
function decreases to $1/\sqrt{e}$. The approximate formula shown
in Eq.~(\ref{C_Eq_5}) can be extended to describe the behavior of
the $f^{(1)}$ function far from the center of the condensate. For
$\delta_L^2(T)\gg1$, the coherence length $L_{\phi}$ is small
compared to the axial size $L$, and the system is well described
by means of the local density approximation
\cite{Dettmer01,Hellweg01,Gerbier02a}. As pointed out in Ref.\
\cite{Gerbier02a}, this limit is equivalent to the use of the
approximate formula for the Jacobi polynomials with large $j$
\cite{Gradsteyn65}. Equation~(\ref{C_Eq_4}) can thus be written in
the form
\begin{equation}
f^{(1)}(z_1/L,z_2/L)=\frac{|z_1-z_2|/L}{[1-(z_1+z_2)^2/(2L)^2]^2},
\label{C_Eq_10}
\end{equation}
generalizing the result obtained in Eq.~(\ref{C_Eq_5}).

\begin{figure}
\includegraphics{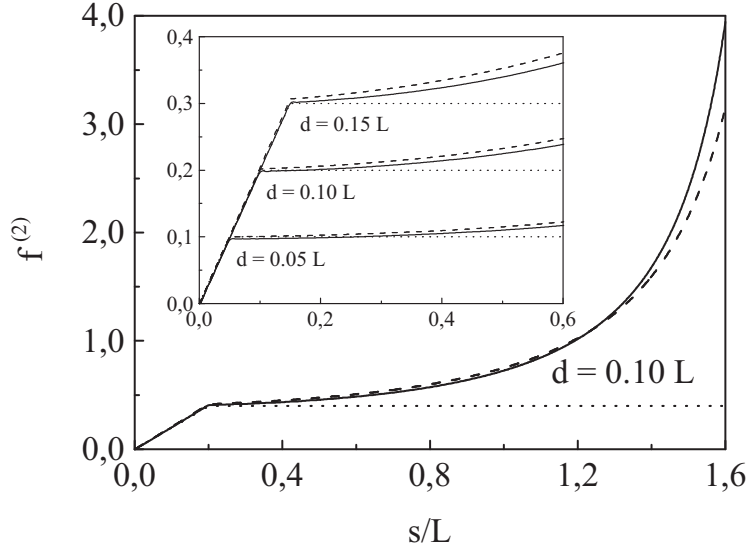}
\caption{\label{C_Fig_1}$f^{(2)}(\bar{z}_1/L,\bar{z}_2/L,\bar{z}_3/L,\bar{z}_4/L)$
as a function of $s>0$. The complete expression in
Eq.~(\ref{C_Eq_15}) (solid line) is compared with the approximated
formulas derived from Eqs.~(\ref{C_Eq_5}) and (\ref{C_Eq_10}),
valid in the condensate center (dotted line) and in the local
density approximation (dashed line). The inset shows $f^{(2)}$ for
different values of $d>0$.}
\end{figure}

We use a similar approach to calculate the two-particle
correlation function of the condensate. Introducing the operator
$\delta^{(2)}\hat{\phi}({\bf r}_1,{\bf r}_2, {\bf r}_3,{\bf r}_4)=
\hat{\phi}({\bf r}_1)+\hat{\phi}({\bf r}_2)- \hat{\phi}({\bf
r}_3)-\hat{\phi}({\bf r}_4)$, we obtain
\begin{equation}
\langle\hat{\psi}^\dagger({\bf r}_1) \hat{\psi}^\dagger({\bf r}_2)
\hat{\psi}({\bf r}_3)\hat{\psi}({\bf r}_4)\rangle=
\prod_{i=1}^4\sqrt{n_0({\bf r}_i)} \exp{\{-\langle[
\delta^{(2)}\hat{\phi}({\bf r}_1,{\bf r}_2,{\bf r}_3, {\bf
r}_4)]^2\rangle/2\}}.
\label{C_Eq_11}
\end{equation}
Using Eq.~(\ref{operphi}) for the phase operator, a
straightforward calculation yields
\begin{eqnarray}
\langle[\delta^{(2)}\hat{\phi}(z_1,z_2,z_3,z_4)]^2\rangle=&&
\sum_{j=1}^{\infty}
\frac{(j+2)(2j+3)\mu}{15(j+1)\epsilon_jN_0}N_j \nonumber\\
&&\times\left[P_j^{(1,1)}\left(\frac{z_1}{L}\right)
+P_j^{(1,1)}\left(\frac{z_2}{L}\right)\right. \nonumber\\
&&\left.-P_j^{(1,1)}\left(\frac{z_3}{L}\right)-
P_j^{(1,1)}\left(\frac{z_4}{L}\right)\right]^2.
\label{C_Eq_12}
\end{eqnarray}
In the limit $k_BT\gg\hbar\omega_z$, the thermal average of
Eq.~(\ref{C_Eq_12}) gives
\begin{equation}
\langle[\delta^{(2)}\hat{\phi}(z_1,z_2,z_3,z_4)]^2\rangle_T=
\delta_L^2(T)f^{(2)}(z_1/L,z_2/L,z_3/L,z_4/L),
\label{C_Eq_13}
\end{equation}
where
\begin{eqnarray}
f^{(2)}(z_1/L,z_2/L,z_3/L,z_4/L)=&&
f^{(1)}(z_1/L,z_3/L)+f^{(1)}(z_2/L,z_4/L) \nonumber \\
&&-f^{(1)}(z_1/L,z_2/L)-f^{(1)}(z_3/L,z_4/L) \nonumber \\
&&+f^{(1)}(z_1/L,z_4/L)+f^{(1)}(z_2/L,z_3/L).
\label{C_Eq_15}
\end{eqnarray}
Thus, the two-particle correlation function can be expressed as a
product of one-particle correlation functions.
Equations~(\ref{C_Eq_5}) and (\ref{C_Eq_10}) can be used to derive
simplified expressions for the $f^{(2)}$ function, valid in the
limit $|z_i|\ll L$ ($i=1,\ldots,4$) and in the local density
approximation. Figure~\ref{C_Fig_1} shows the dependence of
$f^{(2)}$ calculated in
\begin{equation}
\bar{z}_1=\frac{d+s}{2},\quad \bar{z}_2= \frac{-d-s}{2},\quad
\bar{z}_3=\frac{-d+s}{2},\quad \bar{z}_4= \frac{d-s}{2}
\label{C_Eq_16}
\end{equation}
as a function of $s>0$.
The full expression of $f^{(2)}$ can be compared with the two
approximated formulas, the first valid in the condensate center,
the second valid in the local density approximation. The inset of
Fig.~\ref{C_Fig_1} shows the same curves for different values of
$d>0$. This choice of variables follows the particular
experimental realization. In Sec.~IV, we demonstrate how these
curves can be measured in a matter wave interferometry experiment.
There, $d$ is the displacement between the two interfering
condensate copies, and $s$ is the separation between the positions
in the interference pattern at which the particle densities are
evaluated.

A qualitative understanding of the behavior shown in
Fig.~\ref{C_Fig_1} is possible if we consider that
\begin{eqnarray}
\langle[\delta^{(2)}\hat{\phi}(\bar{z}_1,\bar{z}_2,
\bar{z}_3,\bar{z}_4)]^2\rangle_T=&&
\langle[\delta\hat{\phi}(\bar{z}_1,\bar{z}_3)]^2\rangle_T+
\langle[\delta\hat{\phi}(\bar{z}_2,\bar{z}_4)]^2\rangle_T \nonumber \\
&&+2\langle\delta\hat{\phi}(\bar{z}_1,\bar{z}_3)
\delta\hat{\phi}(\bar{z}_2,\bar{z}_4)\rangle_T.
\label{C_Eq_17}
\end{eqnarray}
The first and the second term are the thermal averages of the
operator $(\delta\hat{\phi})^2$ calculated in
$(\bar{z}_1,\bar{z}_3)$ and in $(\bar{z}_2,\bar{z}_4)$; the last
term is proportional to the correlation function of
$\delta\hat{\phi}$ at the same coordinates. For a fixed
displacement $d$, when the examined positions are close to the
condensate center ($d,s \ll L$), the first two terms of
Eq.~(\ref{C_Eq_17}) do not depend on the separation $s$. However,
as $s$ rises from $0$ to $d$, the third term increases from
$-2\langle[\delta\hat{\phi}(\bar{z}_1,\bar{z}_3)]^2\rangle_T$
(complete anticorrelation) to its maximum value $0$, resulting in
an uncorrelated phase difference for every $s\geq d$. In the
interval $0\leq s\leq d$, the $f^{(2)}$ function depends linearly
on $s$ with slope 2.

The second order correlation function is defined as
\begin{equation}
g^{(2)}_T({\bf r}_1,{\bf r}_2,{\bf r}_3,{\bf r}_4)=
\frac{\langle\hat{\psi}^\dagger({\bf r}_1) \hat{\psi}^\dagger({\bf
r}_2) \hat{\psi}({\bf r}_3)\hat{\psi}({\bf r}_4)\rangle_T}
{(\langle\hat{\psi}^\dagger({\bf r}_1) \hat{\psi}({\bf
r}_1)\rangle_T\ldots \langle\hat{\psi}^\dagger({\bf r}_4)
\hat{\psi}({\bf r}_4)\rangle_T)^{1/2}}.
\label{C_Eq_18}
\end{equation}
Substituting Eqs.~(\ref{C_Eq_11}) and (\ref{C_Eq_13}) in
Eq.~(\ref{C_Eq_18}), we obtain:
\begin{equation}
g^{(2)}_T(z_1,z_2,z_3,z_4)=
\exp\{-\delta_L^2(T)f^{(2)}(z_1/L,z_2/L,z_3/L,z_4/L)/2\}.
\label{C_Eq_19}
\end{equation}
Note that, due to the suppression of density modulations, the
normalized density correlation function of the trapped condensate
is constant: $g^{(2)}_T(z_1,z_2,z_2,z_1)=1$.

The calculation we have described for the second order correlation
function can be extended to obtain a general expression for the
$N$-th order correlation function. Defining the operator
\begin{equation}
\delta^{(N)}\hat{\phi}(\{{\bf r}_i\}_{i=1,\ldots,2N})=
\hat{\phi}({\bf r}_1)+\ldots+\hat{\phi}({\bf r}_N)-
\hat{\phi}({\bf r}_{N+1})-\ldots-\hat{\phi}({\bf r}_{2N}),
\label{C_Eq_20}
\end{equation}
the $N$-particle correlation function is given by
\begin{equation}
\langle\hat{\psi}^\dagger({\bf r}_1)\ldots
\hat{\psi}^\dagger({\bf r}_N)
\hat{\psi}({\bf r}_{N+1})\ldots
\hat{\psi}({\bf r}_{2N})\rangle=
\prod_{i=1}^N\sqrt{n_0({\bf r}_i)}
\exp{\{-\langle[\delta^{(N)}\hat{\phi}
(\{{\bf r}_i\}_{i=1,\ldots,2N})]^2
\rangle/2\}}.
\label{C_Eq_21}
\end{equation}
In general, the thermal average of the operator
$(\delta^{(N)}\hat{\phi})^2$ can be written in the form
\begin{equation}
\langle[\delta^{(N)}\hat{\phi}
(\{{\bf r}_i\}_{i=1,\ldots,2N})]^2
\rangle_T=\delta_L^2(T)f^{(N)}
(\{z_i/L\}_{i=1,\ldots,2N}).
\label{C_Eq_22}
\end{equation}
The $f^{(N)}$ function, depending on the Jacobi polynomials
$P_j^{(1,1)}$, can be expressed as a combination of $f^{(1)}$
functions:
\begin{equation}
f^{(N)}(\{z_i/L\}_{i=1,\ldots,2N})=\sum_{1\leq l<m\leq2N}
\mathcal{P}^{\{l,m\}}
f^{(1)}\left(\frac{z_l}{L},\frac{z_m}{L}\right), \label{C_Eq_23}
\end{equation}
where the coefficient $\mathcal{P}^{\{l,m\}}$ is defined as
\begin{equation}
\mathcal{P}^{\{l,m\}}=\left\{
\begin{array}{l}
+1 \quad \textrm{if $l\leq N<m$} \\
-1 \quad \textrm{if $l,m\leq N$ or $l,m> N$}
\end{array}
\right..
\label{C_Eq_24}
\end{equation}
The $N$-th order correlation function is given by
\begin{equation}
g^{(N)}_T(\{{\bf r}_i\}_{i=1,\ldots,2N})=
\frac{\langle\hat{\psi}^\dagger({\bf r}_1)\ldots
\hat{\psi}^\dagger({\bf r}_N) \hat{\psi}({\bf r}_{N+1})\ldots
\hat{\psi}({\bf r}_{2N})\rangle_T}
{(\langle\hat{\psi}^\dagger({\bf r}_1) \hat{\psi}({\bf
r}_1)\rangle_T\ldots \langle\hat{\psi}^\dagger({\bf r}_{2N})
\hat{\psi}({\bf r}_{2N})\rangle_T)^{1/2}}
\label{C_Eq_25}
\end{equation}
and, from Eqs.~(\ref{C_Eq_21}) and (\ref{C_Eq_22}),
\begin{equation}
g^{(N)}_T(\{z_i\}_{i=1,\ldots,2N})=
\exp\{-\delta_L^2(T)f^{(N)}(\{z_i/L\}_{i=1,\ldots,2N})/2\}.
\label{C_Eq_26}
\end{equation}
This general result shows that the spatial correlation function of
phase fluctuating condensates is completely characterized by the
parameter $\delta^2_L(T)$ and, therefore, by the phase coherence
length $L_{\phi}$.
%
\section{Interferometric measurement of the second order correlation function}
The coherence of a matter wave can be studied by using
interferometric methods. However, as standard interference
experiments measure the first order correlation function of the
field operator $\hat{\psi}$, they are very sensitive to phase
noise introduced by the experimental apparatus. The method
presented here is analogous to the original Hanbury-Brown and
Twiss experiment \cite{Hanbury56,Hanbury56a} in which the
spatially resolved second order correlation function $g^{(2)}({\bf
r}_1,{\bf r}_2,{\bf r}_2,{\bf r}_1)$ of a light source is obtained
from intensity correlation measurements. As discussed before, for
a highly elongated BEC $g^{(2)}_T(z_1,z_2,z_2,z_1)=1$. This result
suggests that a simple measurement of density correlations in the
condensate is not sufficient to describe the coherence properties
of the sample. Nevertheless, by measuring density correlations in
the interference pattern generated by two spatially displaced
copies of a parent BEC, it is possible to correlate the field
operator $\hat{\psi}$ at four different positions and extract
$g^{(2)}_T(z_1,z_2,z_3,z_4)$ directly. Compared to standard
interference experiments, the main advantage of this technique is
the intrinsic stability of the density correlation measurement
against variations of the global phase between the interfering
condensates.

In this section, we show how a matter wave Bragg interferometer
can be used to characterize the second order correlation function
of the condensate and measure its phase coherence length.
%
\subsection{Interferometric scheme}
\begin{figure}
\includegraphics[width=0.60\textwidth]{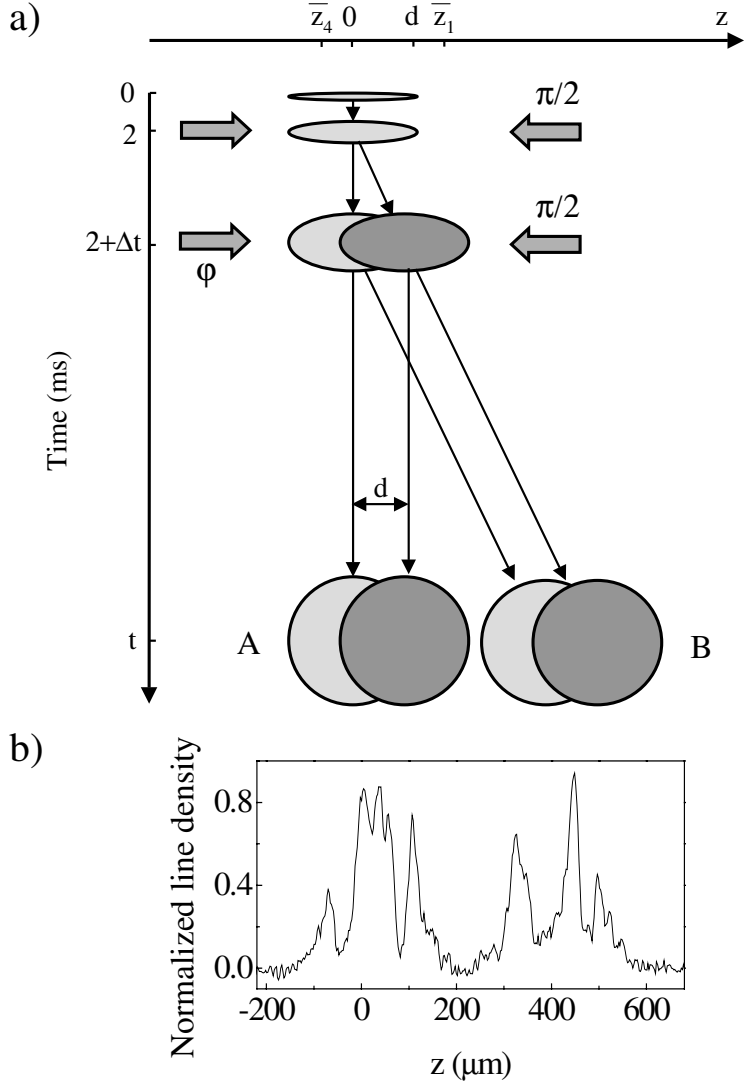}
\caption{\label{I_Fig_1}a) Matter wave Bragg interferometer. The
condensate is released from the magnetic trap and evolves freely
for $2\,\textrm{ms}$. The sample is interrogated by the first
$\pi/2$ Bragg pulse which splits the parent BEC in two copies with
momenta $0$ and $2\hbar k$. After a time $\Delta t$, the second
$\pi/2$ Bragg pulse splits the condensates again and allows them
to interfere. The time interval $\Delta t$ defines the
displacement $d$ between the two interfering condensates. b) A
typical line density profile at the output ports of the
interferometer. The distance between the two autocorrelated copies
($d=46\,\mu\textrm{m}$) is comparable to the phase coherence
length of the parent condensate ($L_{\phi}=43\,\mu\textrm{m}$).In
the schematic of the matter wave Bragg interferometer, the
distance $d$ has been exaggerated for clarity.}
\end{figure}

Our interferometric sequence is shown in Fig.~\ref{I_Fig_1}.
The condensate is released from the magnetic trap and expands
freely for $2\,\textrm{ms}$. This short time-of-flight is
important to lower the density, thus reducing $s$-wave scattering
processes occurring during the Bragg diffraction of the condensate
\cite{Chikkatur00}. The interrogation sequence consists of two
$\pi/2$ Bragg pulses. Each pulse is composed of two
counterpropagating laser beams of wave number $k$, detuned from
the atomic transition. The first Bragg pulse splits the condensate
in the two momentum eigenstates $|2\hbar k\rangle$ and $|0\rangle$
along the axial direction ($z$). After a time $\Delta t$, a second
$\pi/2$ pulse splits the condensates again, creating two
interfering copies in each momentum state. The time interval
$\Delta t$ between the two pulses sets the spatial overlap,
$d=2\hbar k\Delta t/m$, between the interfering BECs at the output
ports of the interferometer. The relative phase of the two
counterpropagating Bragg beams is externally controlled by an
electro-optic modulator ($\textrm{EOM}$) and can be changed
between the two pulses. This allows us to imprint an extra phase
${\varphi}$ which can be precisely tuned. Control of the
$\textrm{EOM}$ phase is crucial for our method, as described in
Sec.~IV-B.

Using the results derived in Sec.~II, the atoms detected in the
output port A (Fig.~\ref{I_Fig_1}), after a total time-of-flight
$t$, are described by the order parameter
\begin{equation}
\psi({\bf r},d,t)=\frac{1}{2}\sqrt{\eta({\bf
r}',t)}+\frac{1}{2}\sqrt{\eta({\bf r},t)}
\exp\{i[\delta\phi(z,z',t)+\alpha(z,z',t)+\beta(z,z')+\gamma(d)]\},
\label{I_Eq_1}
\end{equation}
where ${\bf r}'={\bf r}-d\,{\bf\hat{z}}$ and $\eta({\bf r},t)$ is
the time-evolved density profile normalized to the total number of
atoms in the parent condensate. The relative phase between the
interfering condensates contains several contributions.
$\delta\phi(z,z',t)=\phi(z,t)-\phi(z',t)$ describes the phase
difference between $z$ and $z'$ that evolves from the phase
fluctuations in the parent condensate. The term
\begin{equation}
\alpha(z,z',t)=\frac{m\dot{\lambda}_z}{2\hbar\lambda_z}
(z^2-z'\,^2) \label{I_Eq_2}
\end{equation}
represents the non-uniform spatial phase profile developed during
the mean-field-driven expansion. The mean-field gradient between
the interfering BECs is responsible for a force repelling the
centers of mass of the two clouds. This effect is described by the
phase term
\begin{equation}
\beta(z,z')=\frac{m\delta v}{2\hbar}(z+z'),
\label{I_Eq_3}
\end{equation}
proportional to the relative repulsion velocity $\delta v$ between
the interfering condensates \cite{Simsarian00}. After the first
Bragg pulse the relative phase of the atoms in the $|2\hbar
k\rangle$ momentum state evolves with a characteristic frequency
$\delta_{\mbox{\small Bragg}}$, given by the detuning of the
lasers from the resonance of the two-photon transition
\cite{Weiss94}. Therefore, the last term,
\begin{equation}
\gamma(d)=\delta_{\mbox{\small Bragg}}\Delta t+\varphi=
\delta_{\mbox{\small Bragg}}\frac{m d}{2\hbar k}+\varphi,
\label{I_Eq_4}
\end{equation}
represents a global phase depending on the detuning from the Bragg
transition and the externally controlled phase $\varphi$.

The density of atoms at the output port A of the interferometer is
given by
\begin{eqnarray}
I({\bf r},d,t)=&&\frac{1}{4}\eta({\bf r},t)+
\frac{1}{4}\eta({\bf r}',t) \nonumber\\
&&+\frac{1}{2}\sqrt{\eta({\bf r},t)\eta({\bf r}',t)}
\cos[\delta\phi(z,z',t)+\alpha(z,z',t)+\beta(z,z')+\gamma(d)].
\label{I_Eq_5}
\end{eqnarray}
The presence of strong phase fluctuations alters the interference
pattern generated by the two autocorrelated condensates. In fact,
when $d\simeq L_{\phi}$ the phase term $\delta\phi$ can be
comparable to $\pi$, modifying drastically and in an unpredictable
way the position and the spacings of the interference fringes.
%
\subsection{Method}
Starting from Eq.~(\ref{I_Eq_5}), we want to calculate the density
correlation function of the interference pattern, for an ensemble
of identically prepared condensates at a given temperature $T$,
averaged over all the global phase values $\varphi$. This
averaging process is indicated by the symbol
$\langle\ldots\rangle_{T,\,\varphi}$. It is therefore important
that the phase delay $\varphi$ induced by the $\textrm{EOM}$ is
uniformly changed between $0$ and $2\pi$. In Sec.~II, we have
shown that, for typical times-of-flight (tens of milliseconds),
the evolution of the fluctuating phase of the condensate is
basically frozen. This allows us to neglect the time-dependence of
$\delta\phi(z,z',t)$. We also neglect the contribution of density
modulations induced by the initial phase pattern on the
Thomas-Fermi profile of the condensate. The validity of this
approximation is verified below. Under these assumptions, we
calculate the normalized density correlation function
\begin{equation}
\gamma^{(2)}({\bf r}_1,{\bf r}_2,d,t)= \frac{\langle(I_1-\langle
I_1\rangle_{T,\,\varphi})(I_2-\langle
I_2\rangle_{T,\,\varphi})\rangle_{T,\,\varphi}}
{\sqrt{\langle(I_1-\langle
I_1\rangle_{T,\,\varphi})^2\rangle_{T,\,\varphi}\langle(I_2-
\langle I_2\rangle_{T,\,\varphi})^2\rangle_{T,\,\varphi}}},
\label{M_Eq_1}
\end{equation}
where $I_{1,2}=I({\bf r}_{1,2},d,t)$. After a lengthy but
straightforward calculation, the averaging process gives
\begin{eqnarray}
\gamma^{(2)}(z_1,z_2,d,t)=&&\cos\left[\frac{m}{\hbar}\left(
\frac{\dot{\lambda}_z}{\lambda_z}d+\delta
v\right)(z_1-z_2)\right] \nonumber \\
&&\times\exp[-\delta_L^2(T)f^{(2)}(z_1/L,(z_2-d)/L,(z_1-d)/L,z_2/L)/2].
\label{M_Eq_2}
\end{eqnarray}
$\gamma^{(2)}(z_1,z_2,d,t)$ results from the product of two
different terms: the first is a periodic function, whose argument
is the contribution of the mean-field energy to the phase profile
(ballistic expansion and relative repulsion between the
interfering condensates); the second is an exponential term which
corresponds to the $g^{(2)}_T$ function of the parent phase
fluctuating condensate. The decay constant of this function is
given by the phase coherence length of the condensate (see
Eq.~\ref{C_Eq_9}).

From the experimental point of view, the averaging process
described above is equivalent to the following procedure: The
radially integrated density profile $I=I(z,d,t)$ at the output
port A of the interferometer is measured for different values of
the global phase, uniformly distributed in the range
$0\leq\varphi<2\pi$; then the average value $\langle
I\rangle_{T,\varphi}$ is calculated and used to determine
$I-\langle I\rangle_{T,\varphi}$ for each experimental
realization. These profiles, averaged according to
Eq.~(\ref{M_Eq_1}), give a measurement of
$\gamma^{(2)}(z_1,z_2,d,t)$. We evaluate the density correlations
as a function of the separation $s=z_2-z_1$. For simplicity, we
choose symmetric positions around the center ($z=d/2$) of the
interference pattern in the output port A. The positions in
Eq.~(\ref{C_Eq_16}) are defined such that $z_1=\bar{z}_1$ and
$z_2=\bar{z}_4$ (see Fig.~\ref{I_Fig_1}). The method described
here allows us to characterize the dependence of the correlation
function
\begin{eqnarray}
\gamma^{(2)}(s,d,t)=&&\cos\left[\frac{m}{\hbar}\left(
\frac{\dot{\lambda}_z}{\lambda_z}d+\delta
v\right)s\right] \nonumber \\
&&\times\exp[-\delta_L^2(T)
f^{(2)}(\bar{z}_1/L,\bar{z}_2/L,\bar{z}_3/L,\bar{z}_4/L)/2]
\label{M_Eq_3}
\end{eqnarray}
on the separation $s$ for any fixed displacement $d$ between the
interfering condensates.
%
\subsection{Experimental results and numerical simulations}
We perform the experiment with $^{87}\textrm{Rb}$ condensates in
the $F=1$, $m_F=-1$ state. The atoms are confined in a highly
elongated magnetic trap with cylindrical symmetry, the long axis
lying in the horizontal plane. The confining potential has an
axial frequency $\omega_z=2\pi\times3.4\,\textrm{Hz}$ and a radial
frequency $\omega_{\rho}$ which is varied between
$2\pi\times300\,\textrm{Hz}$ and $2\pi\times380\,\textrm{Hz}$.
Further details on the experimental apparatus can be found in
\cite{Kreutzmann03}. After the BEC formation, we let the system
thermalize in the magnetic trap for typically $4\,\textrm{s}$ in
presence of radio frequency shielding \footnote{We measure the
typical thermalization times of our condensates by performing
experiments of BEC growth. After times of few hundreds of
milliseconds the system reaches an equilibrium condition.}. That
time is important to reach an equilibrium condition in which any
quadrupole oscillation has been damped down. As shown in
Fig.~\ref{I_Fig_1}, our matter wave interferometer consists of two
$\pi/2$ Bragg diffraction pulses. Each of them is composed of two
counterpropagating laser beams, detuned by about $3\,\textrm{GHz}$
from the atomic transition. This detuning suppresses spontaneous
scattering of photons during the interrogation time. The Bragg
pulse duration of $100\,\mu\textrm{s}$ is sufficiently short not
to resolve the momentum distribution of the atoms in the
condensate and long enough to avoid higher order Bragg diffraction
processes. A fixed frequency difference is set between the two
counterpropagating beams to match the Bragg condition. The
condensate is released from the magnetic trap and after
$2\,\textrm{ms}$ of time-of-flight is probed by the two-pulse
sequence of the interferometer. The atomic cloud is detected after
the ballistic expansion by resonant absorption imaging.

Figure~\ref{I_Fig_1}b shows a typical line density profile of an
interference pattern where the distance between the two
autocorrelated copies ($d=46\,\mu\textrm{m}$) is comparable to the
phase coherence length of the parent condensate
($L_{\phi}=43\,\mu\textrm{m}$). Because of the stochastic nature
of phase fluctuations, the fringe spacing is not regular and
differs in each experimental realization. This experimentally
demonstrates that the fluctuating phase of the condensate can
significantly change on distances comparable with the phase
coherence length of the sample. Even if each single image shows
high contrast, the interference pattern is completely washed out
when we average a significant number of realizations.

\begin{figure}
\includegraphics{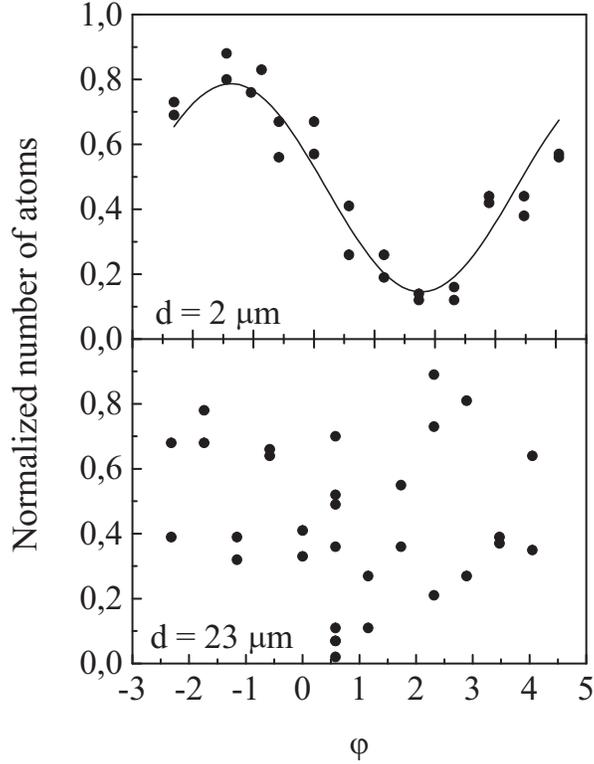}
\caption{\label{E_Fig_1}The number of atoms measured in the
interval $d/2-0.1\times L<z<d/2+0.1\times L$, around the center of
the interference pattern detected at the output port A, is plotted
as a function of the phase $\varphi$ controlled by the
electro-optic modulator. The signal is normalized to the
corresponding number of atoms in the parent condensate. The two
sets of data correspond to different displacements $d$ between the
overlapping condensates. The solid line is obtained by fitting the
experimental data with a sinusoidal function. The measurements
refer to condensates with about $3\times10^5$ atoms, a typical
axial size of $L=180\,\mu\textrm{m}$ and a temperature
$T=170\,\textrm{nK}$ .}
\end{figure}

The results of standard interference experiments are related to
the correlations of the wavefunction and therefore are very
sensitive to phase instabilities. Figure~\ref{E_Fig_1} shows the
interference signal obtained by measuring the number of atoms in
an interval of width $0.2\times L$ around the center of the
interference pattern ($z=d/2$) at the output port $A$, as a
function of the global phase $\varphi$ controlled by the
$\textrm{EOM}$. This signal is normalized to the corresponding
number of atoms in the parent condensate.
The two plots correspond to different displacements $d$ between
the interfering condensates. A small displacement is related to a
short time interval between the two interrogation Bragg pulses. In
that case, the contribution of phase fluctuations and the effect
of technical phase noise introduced by the experimental apparatus
are both negligible. Therefore, according to Eq.~(\ref{I_Eq_5}),
when $d\ll L_{\phi}$ and $\Delta t$ is small compared to the
characteristic time stability of our Bragg pulses, the normalized
signal oscillates sinusoidally with high contrast. For $d$
approaching $L_{\phi}$, the random phase introduced by the phase
fluctuations washes out the oscillation. If external disturbances
can be neglected, the contrast of the oscillations is directly
related to the first order correlation function $g^{(1)}$ at a
given displacement $d$. However, as $d$ increases, the external
disturbances \footnote{For example, a change in the release
velocity of $5.5\times10^{-3}v_{rec}$, where $v_{rec}=\hbar
k/m=5.9\,\mu \textrm{m}/\textrm{ms}$ is the recoil velocity on the
$D_2$ line, leads to a phase change of $\pi/2$ during a time
$\Delta t=3\,\textrm{ms}$ between the Bragg pulses. This value
approximately corresponds to the stability of the release velocity
in our experimental apparatus.} also increase and produce a random
phase noise which destroys the oscillating behavior and hides the
effect of phase fluctuations on the detected signal.

\begin{figure}
\includegraphics{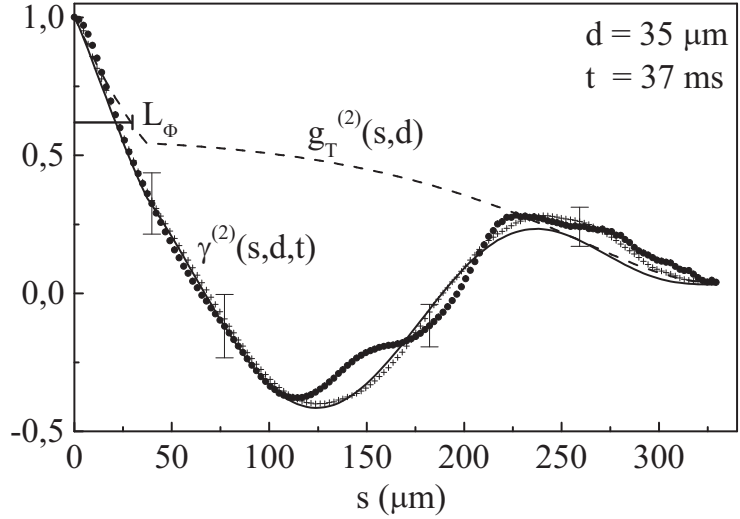}
\caption{\label{E_Fig_2}Circles: Correlation function
$\gamma^2(s,d,t)$ extracted from a set of 29 line density
profiles. The data correspond to samples with $5.0\times10^5$
condensed atoms at a temperature $T=216\,\textrm{nK}$, detected
after a total time-of-flight $t=37\,\textrm{ms}$. The displacement
between the interfering BECs is $d=35\,\mu\textrm{m}$. The bars on
the experimental points represent the statistical errors. Crosses:
Numerical simulation which takes into account the time dependence
of the fluctuating phase and of density modulations, modelled on
the experimental parameters. Solid line: Fit to the experimental
data using the model function of Eq.~(\ref{E_Eq_1}). Dashed line:
Second order correlation function $g^{(2)}_T(s,d)=
g^{(2)}_T(\overline{z}_1,\overline{z}_2,\overline{z}_3,\overline{z}_4)$
extracted from the fit to the experimental data. The phase
coherence length of the sample is graphically indicated on the
plot.}
\end{figure}

\begin{figure}
\includegraphics{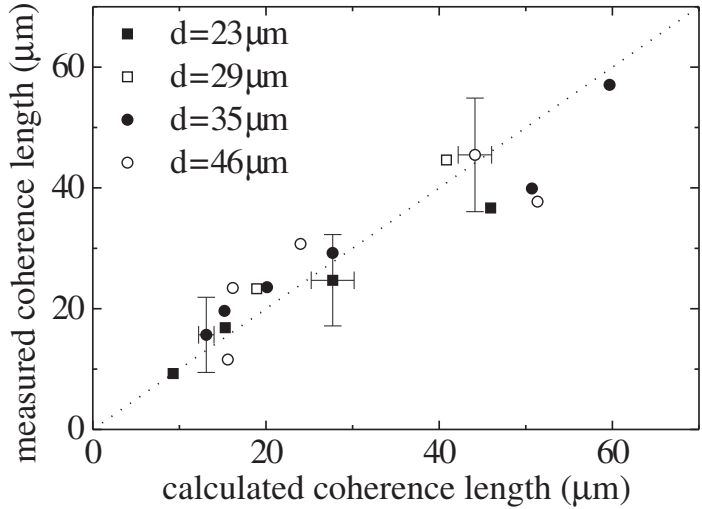}
\caption{\label{E_Fig_3}Direct comparison between the measured
phase coherence lengths and the theoretical values, calculated
according to Eq.~(\ref{C_Eq_9}) by using the measured numbers of
atoms, temperatures and trapping frequencies. The dotted line with
slope 1 is used to compare experiment and theory. The bars on the
plotted points indicate statistical errors. The relative
systematic uncertainties on the calculated and measured phase
coherence length are $26\%$ and $15\%$, respectively. This figure
has previously been shown in \cite{Hellweg03}.}
\end{figure}

This problem can be solved by using the method described in
Sec.~IV-B. The measurement of intensity correlations, in
combination with the subsequent averaging process, has the major
advantage of being insensitive to technical phase noise introduced
by the experimental apparatus. Figure~\ref{E_Fig_2} shows the
correlation function $\gamma^{(2)}(s,d,t)$ extracted from a set of
29 line density profiles corresponding to $5.0\times10^5$
condensed atoms at a temperature $T=216\,\textrm{nK}$, detected
after a total time-of-flight $t=37\,\textrm{ms}$. The displacement
between the interfering BECs is $d=35\,\mu\textrm{m}$.
The experimental data is compared with a numerical simulation
which produces random phase patterns according to the experimental
conditions and uses Eq.~(\ref{I_Eq_1}) to describe the evolution
of the order parameter. The numerically calculated points shown in
Fig.~\ref{E_Fig_2} are obtained by following the same averaging
procedure we have applied to the experimental data. This kind of
analysis includes the time dependence of the fluctuating phase and
of the density modulations induced by the initial phase pattern.
The solid line is the result of a fit to the experimental data.
According to Eqs.~(\ref{M_Eq_3}) and (\ref{C_Eq_16}), the model
function
\begin{equation}
\cos(a\cdot s)\exp[-b\cdot
f^{(2)}(\overline{z}_1/L,\overline{z}_2/L,
\overline{z}_3/L,\overline{z}_4/L)/2] \label{E_Eq_1}
\end{equation}
contains only two free parameters. The curves clearly show the
damped oscillating behavior. The oscillation frequency strictly
defines the parameter $a$, while the damping coefficient gives a
measurement of $\delta_L^2(T)$. From the fitting function it is
possible to extract the spatial dependence of the second order
correlation function
$g^{(2)}_T(s,d)=g^{(2)}_T(\overline{z}_1,\overline{z}_2,\overline{z}_3,\overline{z}_4)$
(see Eqs.~(\ref{C_Eq_19}) and (\ref{M_Eq_3})). The fit on the
experimental data gives a phase coherence length
$L_{\varphi}^{\mbox{\footnotesize exp}}=(57\pm10)\,\mu\textrm{m}$,
compatible with the expected value
$L_{\varphi}^{\mbox{\footnotesize th}}=(58\pm2)\,\mu\textrm{m}$.
The good agreement between the experimental data, the numerical
simulation and the model function of Eq.~(\ref{E_Eq_1})
demonstrates that the free evolution of density modulations and of
the fluctuating phase pattern do not influence the measurement of
the second order correlation function. This result justifies the
use of Eq.~(\ref{M_Eq_2}) to model the experimental data and to
extract the coherence properties of the condensate.
In Fig.~\ref{E_Fig_3}, we show a direct comparison between the
measured phase coherence lengths in the center of the BEC and the
theoretical values calculated according to Eq.~(\ref{C_Eq_9}), by
using the measured numbers of atoms, temperatures and trapping
frequencies. The bars indicate the statistical errors both on the
measured values and on the theoretical predictions.
The dotted line with slope 1 highlights the good quantitative
agreement between experiments and theory.
%
\section{Conclusion}
In this paper, we have studied the coherence properties of phase
fluctuating Bose-Einstein condensates. In highly elongated BECs
the thermal excitation of quasiparticle modes can significantly
reduce the coherence length of the system. Starting from the
results of Petrov et al. \cite{Petrov01}, we have derived a
general formula for the $N$-particle correlation function. The
second order correlation function has been studied in detail and
its limits both around the center of the condensate and in the
local density approximation have been analyzed. In particular, we
have discussed a method to directly characterize the second order
correlation properties of the system. An analytic theory that
describes the free evolution of the condensate phase has been
developed to closely model the measurement process. Using a Bragg
diffraction interferometer, we have measured the density
correlations of the interference pattern generated by two
spatially displaced copies of a parent BEC. This kind of
measurement allows to correlate the field operator $\hat{\psi}$ of
the parent condensate in four different $z$ positions. The
averaging process directly gives the second order correlation
function. The experiment confirms our theoretical predictions and
demonstrates a method to measure the phase coherence length of the
condensate. Compared to usual interference experiments this
technique has the advantage of being insensitive to the global
phase noise introduced by the experimental apparatus. The method
presented here is in direct analogy to the original Hanbury-Brown
and Twiss experiment and demonstrates the possibility of using
density correlation measurements to study the coherence properties
of Bose-Einstein condensates.
\begin{acknowledgments}
We gratefully thank DFG for the support in the
Sonderforschungsbereich 407 as well as the European Union for
support in the RTN network "Preparation and application of
quantum-degenerate cold atomic/molecular gases", contract
HPRN-CT-2000-00125. L.S.~thanks the Alexander von Humboldt
foundation and the ZIP program of the German government for
support.
\end{acknowledgments}

\bibliography{g2}

\end{document}